\documentclass[%
 reprint,
 amsmath,amssymb,
 aps,
]{revtex4-2}

\usepackage{epsfig}
\usepackage[colorlinks,linkcolor=blue,anchorcolor=blue,
            citecolor=blue,urlcolor=blue,
            breaklinks=true]{hyperref}

\usepackage{graphics}
\usepackage{graphicx}  
\usepackage{color}
\usepackage{dcolumn}
\usepackage{bm}
\usepackage{cases}
\usepackage{appendix}
\usepackage[final]{changes}
\usepackage{xcolor}
\usepackage{pgfplots}
\setaddedmarkup{\textcolor{blue}{#1}}
\setdeletedmarkup{\textcolor{red}{\sout{#1}}}

\begin{document}

\title{Geometry-Induced Vacuum Polarization and Mode Shifts in Maxwell-Klein-Gordon Theory}

\author{Li Wang}
 \altaffiliation[Also at ]{Joint Center for Particle, Nuclear Physics and Cosmology, Nanjing 210093, China}
 \email{DG1822035@smail.nju.edu.cn}
 \thanks{Corresponding author}
\author{Jun Wang}
\email{wangj@nju.edu.cn}%
\affiliation{%
 Department of Physics, Nanjing University, Nanjing, 210093, China }%
\author{Yong-Long Wang}
 \email{wangyonglong@lyu.edu.cn}
 \affiliation{School of Physics and Electrical Engineering, Linyi University, Linyi 276000, China}

\date{\today}

\begin{abstract}

Geometric confinement is known to modify single-particle dynamics through effective potentials, yet its imprint on the interacting quantum vacuum remains largely unexplored. In this work, we investigate the Maxwell--Klein--Gordon system constrained to curved surfaces and demonstrate that the geometric potential $\Sigma_{\mathrm{geom}}(\mathbf{r})$ acts as a local renormalization environment. Going beyond standard approximations, we derive an exact analytical framework where extrinsic curvature modifies the scalar loop spectrum, entering the vacuum polarization as a position-dependent mass correction $M^2(\mathbf{r}) \to m^2 + \Sigma_{\mathrm{geom}}(\mathbf{r})$. This induces a finite, gauge-invariant ``geometry-induced running'' of the electromagnetic response. In the long-wavelength regime ($|{\bf Q}|R \ll 1$), we obtain a closed-form expression for the relative frequency shift $\Delta\omega/\omega$, governed by the overlap between the electric energy density and the geometric potential. Applying this formalism to Gaussian bumps, cylindrical shells, and tori, we identify distinct spectral signatures that distinguish these quantum loop corrections from classical geometric optics. Our results suggest that spatial curvature can serve as a tunable knob for ``vacuum engineering,'' offering measurable shifts in high-$Q$ cavities and plasmonic systems relevant to current nanophotonic experiments.

\end{abstract}


\maketitle


\section{Introduction}
The interplay between geometry and quantum fields is a cornerstone of modern physics, manifesting in phenomena ranging from the Hawking radiation of black holes to spectral modifications in curved graphene sheets. In non-gravitational laboratory settings, confining matter fields to curved thin layers or tubular neighborhoods of surfaces generates effective potentials that govern low-dimensional dynamics. The seminal work by da Costa \cite{daCosta1981} and others \cite{JensenKoppe1971,EncinosaEtemadi1998,FerrariCuoghi2008,Ortix2015} established that extrinsic curvature acts as an emergent kinematic potential for single-particle wavefunctions\replaced{. This has}{, with} measurable consequences for electronic bands and scattering. Analogous ideas now permeate guided-wave and photonic platforms, where curvature and morphology induce effective potentials and dispersion control in dielectric and plasmonic waveguides \cite{Marcuse1991,SnyderLove1983,Katsnelson2006}. 

\added{Despite these advances, the effect of geometric confinement on the interacting quantum vacuum requires further investigation.} \deleted{However, a fundamental question remains largely unaddressed: How does this geometric confinement affect the interacting quantum vacuum?} Standard approaches typically treat the geometric potential as a fixed background affecting only real particles (on-shell states). They often neglect how the confinement modifies the spectrum of \textit{virtual} fluctuations (off-shell states) that underpin radiative corrections. \deleted{This gap is critical because, i}\added{I}n quantum field theory, the vacuum is \deleted{not an empty void but }a dynamic medium responsive to external backgrounds \cite{Schwinger1951,ferreiro2021PRD,PhysRevA.109.052435,yadav2025vacuum}. Recent advances in ``vacuum engineering'' have demonstrated that structuring the electromagnetic environment can fundamentally alter quantum emission and entanglement \cite{Rivera2020NatRevPhys,Baranov2018RevModPhys}. If geometry modifies the effective mass of virtual scalars, it must necessarily alter the vacuum polarization loops\replaced{. This is}{, }analogous to how strain engineering in 2D materials creates synthetic gauge fields \cite{lu2014topological,yang2019synthesis,dai2019strain}.

In this work, we address the challenge of accurately describing quantum field effects in confined geometries. Building on the effective field theory framework established in Ref.~\cite{LWang2025}, we demonstrate that the geometric potential $\Sigma_{\mathrm{geom}}$ acts as a local renormalization environment, not merely a boundary condition. This mechanism leads to a finite, gauge-invariant ``geometry-induced running'' of the electromagnetic coupling \cite{Gorbar_2003,Shapiro_2008}. Unlike standard approaches that treat confinement as a fixed background, our formalism captures how curvature locally modifies the vacuum spectrum. This bridges the gap between abstract field theory and observable phenomena\replaced{. It aligns}{, aligning} with recent experimental efforts to simulate curved-spacetime physics in photonic and condensed matter systems \cite{PhysRevA.96.041804,Steinhauer2021Nature,overstreet2022observation}, \replaced{offering}{thereby offering} a new theoretical handle for these analogue gravity platforms.

\replaced{This perspective extends beyond}{This perspective offers a significant advancement over} classical geometric optics or standard Casimir effect calculations. Unlike geometric optics, which deals with ray trajectories, our effect arises from quantum loops. Unlike the global Casimir effect, our correction is local and determined by the specific profile of the extrinsic curvature. In gauge--matter field theories, such one-loop effective actions determine the scale dependence of low-energy effective parameters \cite{PeskinSchroeder1995,weinberg1995quantum,panangaden1981one,BarvinskyVilkovisky1985}. When masses are spatially modulated by geometry, loop integrals acquire modified finite parts, producing effective running at experimentally accessible scales \cite{AppelquistCarazzone1975,CollinsRenorm1984,Weinberg1980}.

Methodologically, we pursue two complementary routes to quantify this effect. First, a background-field and heat-kernel analysis reveals that $\Sigma_{\mathrm{geom}}(\mathbf{r})$ enters loop physics via the modified propagator, with Seeley--DeWitt coefficients controlling local counterterms \cite{DeWitt1975,Vassilevich2003,BirrellDavies1982,ParkerToms2009,vassilevich2005heat}\added{; see also \cite{franchino2023resummed} for recent adaptations to complex boundary conditions}. Second, a scattering phase-shift approach treats $\Sigma_{\mathrm{geom}}$ as a background potential, allowing us to compute the vacuum polarization $\Pi_{\mu\nu}$ and resulting spectral shifts \cite{BordagAdvPhys2005,RahiPRD2009,EmigPRL2007,MiltonBook2001}\added{, a formalism recently applied to complex multiparticle systems \cite{PhysRevA.108.032802}}.

Our main finding is that the geometric potential imparts a controlled dependence on the curvature radius $R$ to the finite part of $\Pi_{\mu\nu}$. This manifests in observables---effective permittivity, modal frequencies, and group delays---as a geometry-induced shift beyond simple path-length corrections. These predictions are directly testable within relevant platforms. High-$Q$ microwave/optical cavities and bent waveguides allow precision tracking of modal frequencies \cite{HarocheRaimond2006,KippenbergVahala2008,VahalaNature2003}. Plasmonic and metamaterial systems provide tunable curvature and nanoscale morphology, together with established frameworks for nonlocal response \cite{MaierBook2007,CiraciScience2012,MortensenNatComm2014,GarciaDeAbajoACSPho2014}. Superconducting and polaritonic platforms offer complementary metrology for group delay \cite{BaronePaterno1982,CarusottoCiutiRMP2013}.

Furthermore, advances in analogue-gravity platforms justify treating the ambient spacetime as a controllable design variable in our model. In Bose--Einstein condensates, transonic flows realize horizon kinematics \cite{Steinhauer2021Nature,Steinhauer2016NatPhys,Garay2000PRL,Weinfurtner2011PRL,Boada2012PRL,Celi2014PRL}. Transformation optics systematically maps target metrics to spatially varying constitutive parameters \cite{Leonhardt2006Science,Pendry2006Science,Schurig2006Science,Ginis2016Optica,Rechtsman2013Nature}. Meanwhile, strain and curvature in Dirac materials emulate pseudo-gravitational couplings \cite{Levy2010Science,Guinea2010NatPhys,Unruh1981PRL,Philbin2008Science,UnruhSchuetzhold2003PRD}. \replaced{Together, these developments support extending our ambient spaces beyond Euclidean, such as to Ricci-curved environments. This allows the geometric potential }{Together, these developments support extending our ambient spaces beyond Euclidean---e.g., to Ricci-curved environments---so that the geometric potential }$\Sigma_{\mathrm{geom}}$\replaced{ to be engineered and its loop-level finite corrections probed in situ.}{ can be engineered and its loop-level finite corrections probed in situ.}

The remainder of this work is organized as follows. Section~II defines the MKG model incorporating the geometric potential and outlines the calculation of the vacuum polarization. Section~III presents our main results: scaling of the modal frequency shift $\Delta\omega/\omega$. Section~IV applies this framework to representative geometries (Gaussian bumps, cylinders, tori). Section~V summarizes the findings and discusses the broader perspective of geometry as a renormalization environment.

\section{MKG framework with geometric potential}
\label{sec:MKG-geom}

In our previous work~\cite{LWang2025}, we established a general effective field theory framework for scalar fields constrained to curved submanifolds embedded in higher-dimensional Riemannian manifolds. We derived a universal geometric potential, $\Sigma_{\mathrm{geom}}$, which encodes the extrinsic curvature effects and governs the dynamics of the constrained field.

In the present study, we apply this formalism to a $(3+1)$-dimensional quantum field theory context. \replaced{Specifically, we consider the physically relevant scenario where the ambient spacetime is $\mathcal{M}^4 \cong \mathbb{R}^{3,1}$ (or potentially $\mathcal{M}^4 \times \mathcal{K}$ with compact $\mathcal{K}$), and the scalar field is localized to a spatial submanifold, denoted as $\mathcal{N}$.}{Specifically, we consider the physically relevant scenario where the ambient spacetime is $\mathcal{M}^4 \cong \mathbb{R}^{3,1}$ (or potentially $\mathcal{M}^4 \times \mathcal{K}$ with compact $\mathcal{K}$), and the scalar field is localized to a submanifold.}

\replaced{To calculate the vacuum polarization, we treat the field as a fluctuation in the ambient spacetime using the established "physical thin-layer" approach \cite{JensenKoppe1971,daCosta1981}. Specifically, we utilize the effective field theory framework derived in Ref.~\cite{LWang2025}}{While the effective field equation derived in Ref.~\cite{LWang2025} describes the dynamics on the submanifold itself, the calculation of vacuum polarization requires treating the field as a fluctuation in the ambient spacetime. We adopt the "physical thin-layer" approach}, where the geometric potential $\Sigma_{\mathrm{geom}}$ acts as a constitutive mass correction within a layer of finite thickness $h$.

Crucially, we perform the loop integrals over the momenta of the non-compact ambient spacetime ($D=3+1$). We do not integrate over the internal geometric degrees of freedom of the submanifold (such as those of $\mathbb{C}P^1$ discussed later), as their contribution is already effectively captured by the geometric potential $\Sigma_{\mathrm{geom}}$ via the thin-layer quantization procedure.

\replaced{To evaluate the electromagnetic response of this geometric vacuum, we introduce the $U(1)$ gauge field via minimal coupling \cite{PeskinSchroeder1995}. The effective Lagrangian density in the bulk is thus defined as:}{The Lagrangian density in the bulk is thus defined as:}
\begin{equation}
\mathcal{L}
= -\tfrac{1}{4}\,F_{\mu\nu}F^{\mu\nu}
+ (D_\mu\phi)^\ast(D^\mu\phi)
- \big[m^2+\Sigma_{\mathrm{geom}}(\mathbf{r})\big]\,|\phi|^2,
\label{eq:L}
\end{equation}
with
\begin{equation}
D_\mu=\partial_\mu+i q A_\mu,
\qquad
F_{\mu\nu}=\partial_\mu A_\nu-\partial_\nu A_\mu.
\label{eq:defs}
\end{equation}
The curvature-induced potential is assembled from embedding invariants\added{\cite{LWang2025}}:
\begin{equation}
\Sigma_{\mathrm{geom}}(\mathbf{r})
=
\frac{1}{4}\,\|H\|^2
-\frac{1}{2}\,\|\mathrm{II}\|^2
-\frac{1}{2}\sum_{a=1}^{m-n}\mathrm{Ric}_M(\nu_a,\nu_a),
\label{eq:sigma}
\end{equation}
where $H$ is the mean curvature vector, $\mathrm{II}$ the second fundamental form, $\{\nu_a\}$ an orthonormal basis of normal directions to \added{submanifold} $\mathcal{N}$, and $\mathrm{Ric}_M(\nu_a,\nu_a)$ the ambient Ricci curvature projected along those normals. Intuitively, $\Sigma_{\mathrm{geom}}$ is a controlled, slowly varying deformation of the local mass squared, whose effects can be organized perturbatively in amplitude and gradients when $mR\gg 1$ and $\|\Sigma_{\mathrm{geom}}\|\ll m^2$. 

Throughout this work, we adopt natural units ($\hbar = c = 1$) for intermediate field-theoretic derivations to maintain notational clarity. However, to facilitate comparison with spectroscopic measurements, we explicitly restore SI units (including the vacuum permittivity $\varepsilon_0$) in the final expressions for susceptibilities and frequency shifts.

\subsection*{Scalar resolvent and local expansions}
\replaced{Following standard Green's function techniques in curved backgrounds \cite{BirrellDavies1982,ParkerToms2009}, we introduce the static-frequency scalar resolvent}{Introduce the static-frequency scalar resolvent}
\begin{equation}
G_{\Sigma}(\omega)
=\Big[-\nabla^2 - \omega^2 + m^2 + \Sigma_{\mathrm{geom}}(\mathbf{r})\Big]^{-1},
\label{eq:Gsigma}
\end{equation}
with $G_0(\omega)=[-\nabla^2-\omega^2+m^2]^{-1}$. For smooth, weak $\Sigma_{\mathrm{geom}}$ one has the Born/Dyson series
\begin{equation}
G_{\Sigma}
= G_0
- G_0\,\Sigma_{\mathrm{geom}}\,G_0
+ G_0\,\Sigma_{\mathrm{geom}}\,G_0\,\Sigma_{\mathrm{geom}}\,G_0
- \cdots,
\label{eq:Gseries}
\end{equation}
which organizes nonlocal effects as a convolution expansion in $\Sigma_{\mathrm{geom}}$. When $mR\gg 1$, the heat kernel at coincident points admits a controlled local expansion with $M^2(\mathbf{r})\equiv m^2+\Sigma_{\mathrm{geom}}(\mathbf{r})$:
\begin{equation}
\begin{aligned}
K(\mathbf{r},\mathbf{r};s)
&= \frac{e^{-s M^2(\mathbf{r})}}{(4\pi s)^{d/2}}
\Bigg[
1
+ s\,\frac{\nabla^2 M^2}{6\,M^2}
\\
&\quad
+ s^2\,\mathcal{O}\!\left(\frac{\|\nabla M^2\|^2}{M^4}\right)
+ \cdots
\Bigg],
\end{aligned}
\label{eq:HK}
\end{equation}
valid for $d=3$ under the condition that gradients of $M^2$ are small compared to $M^2$ itself. This DeWitt–Seeley–Gilkey series systematically captures local and gradient corrections.

\subsection*{One-loop vacuum polarization with curvature}
\replaced{Using standard quantum field theory methods \cite{PeskinSchroeder1995,weinberg1995quantum}, the connected one-loop photon self-energy reads}{The connected one-loop photon self-energy reads}
\begin{equation}
\Pi_{\mu\nu}(x,x';\Sigma_{\mathrm{geom}})
= i q^2\,
\Big[
G_{\Sigma}(x,x')\,\overleftrightarrow{\partial_\mu}\,
\overleftrightarrow{\partial'_\nu}\,G_{\Sigma}(x',x)
\Big]_{\text{conn}},
\label{eq:Pi-xx}
\end{equation}

with $f\,\overleftrightarrow{\partial_\mu}g \equiv f(\partial_\mu g)-(\partial_\mu f)g$ and the primed derivative acting on $x'$. \added{Note that while the four-point ``seagull'' diagram ($|\phi|^2 A^2$) also contributes to the total self-energy, its loop integral is purely local and independent of the external momentum. It serves only to cancel the longitudinal part of the bubble graph to preserve the Ward identity \cite{PeskinSchroeder1995}, leaving the momentum-dependent transverse form factor $\delta\Pi_T$ unaffected. Therefore, we focus on the connected bubble graph.}

Expanding Eq.~\eqref{eq:Pi-xx} to first order in $\Sigma_{\mathrm{geom}}$, \added{and introducing the external photon four-momentum $Q^\mu = (\Omega, \mathbf{Q})$ (with frequency $\Omega$ and spatial momentum $\mathbf{Q}$), we obtain:}
\begin{align}
&\Pi_{\mu\nu}(\Omega;\mathbf{r},\mathbf{r}')
= \Pi^{(0)}_{\mu\nu}(\Omega;\mathbf{r}-\mathbf{r}')
+ \delta\Pi_{\mu\nu}(\Omega;\mathbf{r},\mathbf{r}')
+ \mathcal{O}(\Sigma_{\mathrm{geom}}^2),
\label{eq:Pi-split}\\
&\delta\Pi_{\mu\nu}
= - i q^2\,
\Big[
G_0\,\Sigma_{\mathrm{geom}}\,G_0\,\overleftrightarrow{\partial_\mu}\,
\overleftrightarrow{\partial'_\nu}\,G_0
+ G_0\,\overleftrightarrow{\partial_\mu}\,
\overleftrightarrow{\partial'_\nu}\,G_0\,\Sigma_{\mathrm{geom}}\,G_0
\Big].
\label{eq:dPi-real}
\end{align}
In momentum space, with external momentum $Q^\mu=(\Omega,\mathbf{Q})$ and $G_0(k)=\big[\mathbf{k}^2+m^2-\Omega^2+i0^+\big]^{-1}$ for the spatial loop momentum $\mathbf{k}$, one obtains
\begin{align}
&\delta\Pi_{\mu\nu}(\Omega;\mathbf{Q})
= q^2
\int \frac{d^3k}{(2\pi)^3}\,
\Big[(2k_\mu{+}Q_\mu)\,(2k_\nu{+}Q_\nu)\Big]\,
\notag\\
&\times
G_0(k)\,G_0(k{+}Q)
\Big[
G_0(k)\,\Sigma_{\mathrm{geom}}(\mathbf{Q})
+ G_0(k{+}Q)\,\Sigma_{\mathrm{geom}}(-\mathbf{Q})
\Big],
\label{eq:dPi-k}
\end{align}
which preserves transversality by the Ward identity. To $\mathcal{O}(\Sigma_{\mathrm{geom}})$ one may write
\begin{equation}
\begin{aligned}
\delta\Pi_{\mu\nu}
&= \big(Q_\mu Q_\nu - Q^2 g_{\mu\nu}\big)\,
\delta\Pi_T(\Omega;\mathbf{Q})
\\
&\quad+ \mathcal{O}(\Sigma_{\mathrm{geom}}^2),
\end{aligned}
\label{eq:transv}
\end{equation}
thereby defining the transverse scalar form factor $\delta\Pi_T$.

\subsection*{Long-wavelength and local limits}
For slowly varying curvature, $|\mathbf{Q}|R\ll 1$, a gradient expansion of $\delta\Pi_T$ yields
\begin{equation}
\begin{aligned}
\delta\Pi_T(\Omega;\mathbf{Q})
&\simeq
\Sigma_{\mathrm{geom}}(\mathbf{0})\,\mathcal{I}_0(\Omega;m)
\\
&\quad
- \tfrac{1}{2}\,Q_i Q_j\,
\overline{\partial_i\partial_j \Sigma_{\mathrm{geom}}}\,
\mathcal{I}_2(\Omega;m)
+ \cdots,
\end{aligned}
\label{eq:dPiT-grad}
\end{equation}
where overlines denote local values in the gradient expansion. The loop integrals are
\begin{align}
\mathcal{I}_0(\Omega;m)
&= q^2 \int\!\frac{d^3k}{(2\pi)^3}\,
\frac{4\mathbf{k}^2 - 3 \Omega^2}
{\big[\mathbf{k}^2+m^2-\Omega^2+i0^+\big]^3},
\label{eq:I0}\\
\mathcal{I}_2(\Omega;m)
&= q^2 \int\!\frac{d^3k}{(2\pi)^3}\,
\frac{\alpha(\mathbf{k},\Omega)}
{\big[\mathbf{k}^2+m^2-\Omega^2+i0^+\big]^4},
\label{eq:I2}
\end{align}
with $\alpha$ a polynomial arising from the $Q$-expansion (its explicit form is unnecessary for the $\mathcal{O}(Q^0)$ local result). Power counting indicates UV convergence in $d=3$ at the shown powers if a gauge-invariant regulator is employed. Substituting Eq.~\eqref{eq:sigma} into Eq.~\eqref{eq:dPiT-grad} leads to the local curvature-induced correction
\begin{equation}
\begin{aligned}
\delta\Pi_T(\Omega;\mathbf{Q}\!\to\!0)
=
\Big(
\tfrac{1}{4}\|H\|^2
- \tfrac{1}{2}\|\mathrm{II}\|^2
- \tfrac{1}{2}\!\sum_a \mathrm{Ric}_M(\nu_a,\nu_a)
\Big)\,
\mathcal{I}_0(\Omega;m) \\
+ \mathcal{O}(\nabla^2\Sigma_{\mathrm{geom}}),
\end{aligned}
\label{eq:dPiT-local}
\end{equation}
which states that, at leading order in gradients, extrinsic and ambient curvatures renormalize the transverse photon self-energy through a local multiplier set by $\Sigma_{\mathrm{geom}}(\mathbf{r})$; nonlocality enters through the $\mathcal{O}(\nabla^2\Sigma_{\mathrm{geom}})$ remainder and higher Seeley–DeWitt terms.

\subsection*{Low-frequency limit and decoupling}
In the static/low-frequency regime $|\Omega|\ll m$, one finds
\begin{equation}
\mathcal{I}_0(\Omega;m)
= \frac{3q^2}{8\pi\,m}\,
\Bigg[1 + \mathcal{O}\!\Big(\frac{\Omega^2}{m^2}\Big)\Bigg],
\label{eq:I0low}
\end{equation}
consistent with heavy-field decoupling: the contribution of the scalar loop is suppressed as $1/m$, with higher powers in $\Omega^2/m^2$ encoding mild dispersion. Combined with the gradient hierarchy justified by $mR\gg 1$, this result underpins the controlled locality of the curvature correction in the deep subgap regime.

\subsection*{Local susceptibility and response}
Using the transverse constitutive relation $\Pi_{ij}(\Omega;\mathbf{r},\mathbf{r}')\simeq -\Omega^2\,\varepsilon_0\,\chi_{ij}(\Omega;\mathbf{r})\,\delta(\mathbf{r}-\mathbf{r}')$, the curvature-induced shift of the transverse susceptibility becomes
\begin{equation}
\begin{aligned}
\delta\chi_T(\Omega;\mathbf{r})
&= \frac{\mathcal{I}_0(\Omega;m)}{\varepsilon_0\,\Omega^2}\,
\Bigg(
\frac{1}{4}\,\|H\|^2
- \frac{1}{2}\,\|\mathrm{II}\|^2\\
&\quad
- \frac{1}{2}\sum_a \mathrm{Ric}_M(\nu_a,\nu_a)
\Bigg)+ \mathcal{O}(\nabla^2\Sigma_{\mathrm{geom}}).
\end{aligned}
\label{eq:dchi}
\end{equation}
In the low-frequency limit $|\Omega|\ll m$:
\begin{equation}
\delta\chi_T(\Omega\!\ll\! m;\mathbf{r})
\simeq
\frac{3q^2}{8\pi\,\varepsilon_0\,m\,\Omega^2}\,
\Bigg(
\frac{1}{4}\,\|H\|^2
- \frac{1}{2}\,\|\mathrm{II}\|^2
- \frac{1}{2}\sum_a \mathrm{Ric}_M(\nu_a,\nu_a)
\Bigg).
\label{eq:dchilow}
\end{equation}
The $\Omega^{-2}$ scaling follows from $\Pi_{ij}\propto -\Omega^2\chi_{ij}$ by definition. Physically, at fixed curvature the scalar loop with effective mass $m$ dresses the photon propagator, enhancing the susceptibility with a universal prefactor $3q^2/(8\pi m)$ that exhibits decoupling. The gradient remainder $\mathcal{O}(\nabla^2\Sigma_{\mathrm{geom}})$ parametrizes nonlocality induced by spatial curvature variations and is suppressed for $|\mathbf{Q}|R\ll 1$ and $R\ll L_{\sigma}$.

\subsection*{Mode frequency shifts (wrapped form)}
For a normalized unperturbed cavity mode $n$ with frequency $\omega_n^{(0)}$ and stored energy $U_n$, the curvature-induced transverse self-energy correction in the long-wavelength limit yields
\begin{widetext}
\begin{align}
\frac{\Delta\omega_n}{\omega_n^{(0)}}
&= \frac{1}{2 U_n}  %
\int d^3 r\,
\frac{|\mathbf{E}_n(\mathbf{r})|^2}
{\varepsilon_0\,\omega_n^{(0)\,2}}\,
\delta\Pi_T\big(\Omega=\omega_n^{(0)};\mathbf{Q}\!\to\!0\big)
\nonumber\\
&= \frac{\mathcal{I}_0(\omega_n^{(0)};m)} %
{2\,\varepsilon_0\,\omega_n^{(0)\,2}\,U_n}
\int d^3 r\, |\mathbf{E}_n(\mathbf{r})|^2\,
\Bigg(
\frac{1}{4}\,\|H\|^2
- \frac{1}{2}\,\|\mathrm{II}\|^2
- \frac{1}{2}\sum_a \mathrm{Ric}_M(\nu_a,\nu_a)
\Bigg)
+ \mathcal{O}(\nabla^2\Sigma_{\mathrm{geom}},\,\Sigma_{\mathrm{geom}}^2).
\label{eq:dw}
\end{align}

\end{widetext}
This representation makes explicit that the frequency correction is the electric-energy–weighted spatial average of curvature invariants on $\mathcal{N}$, modulated by the universal loop factor $\mathcal{I}_0/m$. Concentration of field energy in regions with large $|\kappa_1-\kappa_2|$ (see below) drives the integral more negative, lowering the resonance; positive contributions from mean curvature and ambient Ricci components can partially compensate, depending on geometry.

\subsection*{Flat ambient manifold: a concise illustration}
In a flat ambient space $\mathbb{R}^3$, the ambient Ricci tensor vanishes. For a surface with principal curvatures $\kappa_1,\kappa_2$,
\begin{equation}
\|H\|^2=(\kappa_1+\kappa_2)^2,
\qquad
\|\mathrm{II}\|^2=\kappa_1^2+\kappa_2^2,
\label{eq:HII}
\end{equation}
so the geometric potential reduces to
\begin{equation}
\Sigma_{\mathrm{geom}}
= \frac{1}{4}(\kappa_1+\kappa_2)^2
- \frac{1}{2}(\kappa_1^2+\kappa_2^2)
= -\frac{1}{4}(\kappa_1-\kappa_2)^2 \le 0.
\label{eq:sigmaR3}
\end{equation}
Two immediate consequences follow. First, near-umbilic surfaces ($\kappa_1\approx\kappa_2$) yield a weak geometric potential and hence smaller shifts. Second, highly anisotropic curvature ($|\kappa_1-\kappa_2|$ large), as near saddle-like regions, enhances $|\Sigma_{\mathrm{geom}}|$ and typically increases the magnitude of the susceptibility correction in Eq.~\eqref{eq:dchilow}, leading to more pronounced downshifts of cavity modes via Eq.~\eqref{eq:dw}, provided $mR\gg 1$ and $|\mathbf{Q}|R\ll 1$ hold.

\subsection*{Dimensionality, Regularization, and Renormalization}
The field theory is defined in $D=3+1$ spacetime dimensions. To handle the ultraviolet (UV) divergences inherent in the one-loop vacuum polarization and to strictly preserve gauge invariance (Ward identities), we employ \textit{Dimensional Regularization} (DimReg), analytically continuing the spacetime dimension to $D = 4 - 2\epsilon$.

The total vacuum polarization is split into the flat-space vacuum contribution and the geometry-induced correction: $\Pi_{\mu\nu} = \Pi_{\mu\nu}^{\mathrm{vac}} + \delta\Pi_{\mu\nu}$. The flat-space term $\Pi_{\mu\nu}^{\mathrm{vac}}$ contains the standard UV divergence (pole in $1/\epsilon$), which is absorbed into the field strength renormalization counterterm $Z_3$ (charge renormalization) using the on-shell renormalization scheme, defined by the condition $\Pi(Q^2=0) = 0$.

Focusing on the geometry-induced correction $\delta\Pi_{\mu\nu}$ [Eq.~\eqref{eq:dPi-real}], the insertion of the geometric potential $\Sigma_{\mathrm{geom}}$ effectively acts as a mass vertex. Power counting in $D=4$ reveals that while the bare bubble diagram is quadratically divergent, the diagram with a single mass insertion (corresponding to $\delta\Pi_{\mu\nu}$) behaves as $\int d^4k \, k^2 / (k^2)^3$, which is logarithmically divergent. However, the gauge-invariant transverse projection $\delta\Pi_T$ extracts the finite part of this response.

Specifically, after performing the Wick rotation and the $k_0$ contour integration, the remaining spatial integrals reduce to $d=3$ momentum integrals. The leading local term $\mathcal{I}_0(\Omega;m)$ in Eq.~\eqref{eq:I0} scales asymptotically as $\int^\Lambda k^2 dk \cdot k^{-6} \sim \text{finite}$, rendering the geometry-induced shift UV finite without the need for additional geometric counterterms at this order. Scheme-dependent contact terms that might arise in a hard-cutoff regularization are automatically discarded by DimReg, ensuring that the resulting response satisfies $Q_\mu \delta\Pi^{\mu\nu} = 0$. 

It is important to distinguish the physical regime of our calculation from strictly two-dimensional effective theories. Our result, characterized by the bulk coefficient $\mathcal{I}_0 \propto 3/(8\pi m)$, applies to the \textit{physical thin-layer regime} where the confining potential is strong enough to localize real low-energy modes, yet the layer thickness $h$ allows virtual fluctuations to probe the ambient 3D volume (i.e., the UV cutoff $\Lambda \gg 1/h$). In the opposite limit of \textit{infinite confinement} (freezing out all transverse excitations), the vacuum polarization would be renormalized by 2D counterterms, yielding a significantly smaller coefficient. We focus on the former, as it represents the realistic scenario for finite-width nanostructures embedded in vacuum.


\section{Applications to Representative Geometries}
\label{sec:applications}

In this section, we apply the general formalism to three distinct geometric configurations: a Gaussian bump, a cylindrical shell, and a torus. For each case, we explicitly construct the geometric potential from the embedding properties and analyze the resulting mode-dependent frequency shifts.

\subsection{Ambient Space Models}
Before specifying the submanifold geometry, we must define the ambient manifold $\mathcal{M}$. We contrast two cases:
\begin{itemize}
    \item \textbf{Euclidean Ambient ($\mathbb{R}^3$):} The ambient Riemann tensor vanishes. The geometric potential is determined solely by the extrinsic curvature (second fundamental form $\mathrm{II}$):
    \begin{equation}
    \Sigma_{\mathrm{geom}}^{E}  = -\frac{1}{4}(\kappa_1 - \kappa_2)^2,
    \end{equation}
    where $\kappa_{1,2}$ are the principal curvatures. Note that $\Sigma_{\mathrm{geom}}^{E} \le 0$ always, acting as an attractive potential well.
    \item \textbf{$\mathbb{C}P^1$ Ambient:} This space possesses a constant positive holomorphic sectional curvature. Projecting onto the normal directions, this contributes a constant offset to the potential:
    \begin{equation}
    \Sigma_{\mathrm{geom}}^{CP^1} = \Sigma_{\mathrm{geom}}^{E} - \frac{1}{r_0^2},
    \end{equation}
    where $r_0$ is the characteristic radius of the ambient space. This term leads to a global, geometry-independent shift in the vacuum energy.
\end{itemize}

Physically, this model serves as a prototype for compactified extra dimensions or topological defects in spinor condensates, providing a controlled contrast to the asymptotically flat Euclidean case.

\subsection{Weak Corrugations: The Gaussian Bump}
We first consider a local deformation of a plane, described by the height function $z(\rho) = h \exp(-\rho^2/2\sigma^2)$ in cylindrical coordinates $(\rho, \phi, z)$. This models a "bump" or "dent" defect on a surface. This regime of "weak corrugation" ($h \ll \sigma$) allows us to treat the curvature as a perturbative correction.

\subsubsection{Geometric Potential Construction}
The radial curvature $\kappa_\rho$ and azimuthal curvature $\kappa_\phi$ are approximated by:
\begin{align}
\kappa_\rho &\approx \partial_\rho^2 z = -\frac{h}{\sigma^2}\left(1 - \frac{\rho^2}{\sigma^2}\right) e^{-\rho^2/2\sigma^2}, \\
\kappa_\phi &\approx \frac{1}{\rho}\partial_\rho z = -\frac{h}{\sigma^2} e^{-\rho^2/2\sigma^2}.
\end{align}
The Euclidean geometric potential is proportional to the square of their difference (the anisotropy):
\begin{equation}
\Sigma_{\mathrm{geom}}^{\mathrm{bump}}(\rho)
= -\frac{1}{4}(\kappa_\rho - \kappa_\phi)^2
= -\frac{h^2}{4\sigma^4} \left( \frac{\rho}{\sigma} \right)^4 e^{-\rho^2/\sigma^2}.  
\label{eq:sigma-bump}
\end{equation}
This potential vanishes at the origin ($\rho=0$) where the surface is locally isotropic (umbilic point), and peaks at a characteristic radius $\rho = \sqrt{2}\sigma$.
\begin{figure}[htbp]
    \centering
    \includegraphics[width=0.9\linewidth]{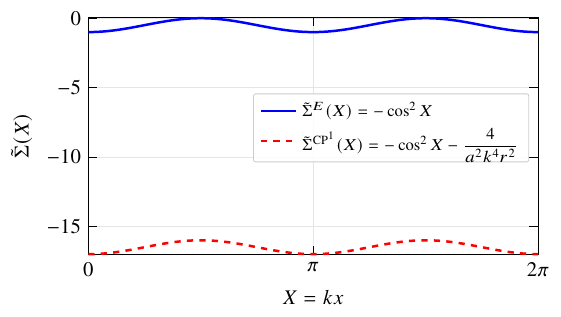} 
    \caption{\textbf{Geometric potential profile and mode selectivity.} 
    (a) The radial dependence of the geometric potential $\Sigma_{\mathrm{geom}}$ for a Gaussian bump. Note that the potential vanishes at the isotropic center ($\rho=0$) and peaks at the region of maximum anisotropy ($\rho \approx \sqrt{2}\sigma$). 
    (b) The overlap between the geometric potential and different electromagnetic modes. Higher-order radial modes (red dashed line) sample the potential peak more effectively than the fundamental Gaussian mode (blue solid line), leading to a larger frequency shift. This demonstrates the mechanism of geometry-induced mode splitting.}
    \label{fig:bump-profile}
\end{figure}

\subsubsection{Frequency Shift and Mode Selectivity}
Substituting Eq.~\eqref{eq:sigma-bump} into the master formula, the relative frequency shift is:
\begin{equation}
\frac{\Delta\omega}{\omega}
= - \frac{\mathcal{I}_0(\omega;m)}{2\varepsilon_0\omega^2} \, F_E \, \frac{h^2}{4\sigma^4} \, C_b,
\end{equation}
where $C_b = \langle (\rho/\sigma)^4 e^{-\rho^2/\sigma^2} \rangle_E$ is the modal overlap factor.$F_E = \varepsilon_0 \int d^3r\, |\mathbf{E}_n(\mathbf{r})|^2 / (2 U_n)$ 
is the electric energy filling fraction of mode $n$ 
($F_E = 1/2$ for a lossless vacuum cavity).

\textbf{Physics Discussion:} The factor $(\rho/\sigma)^4$ introduces a strong \textit{spatial filtering effect}.
\begin{itemize}
    \item For fundamental modes (Gaussian-like) centered at $\rho=0$, the overlap is suppressed because the geometric potential is zero at the center.
    \item For higher-order radial modes (e.g., Laguerre-Gaussian modes) that have intensity maxima near $\rho \approx \sqrt{2}\sigma$, the overlap is maximized.
\end{itemize}
This implies that the geometric vacuum polarization acts as a mode-selective filter.

\begin{figure}[htbp]
\centering
\includegraphics[width=0.9\linewidth]{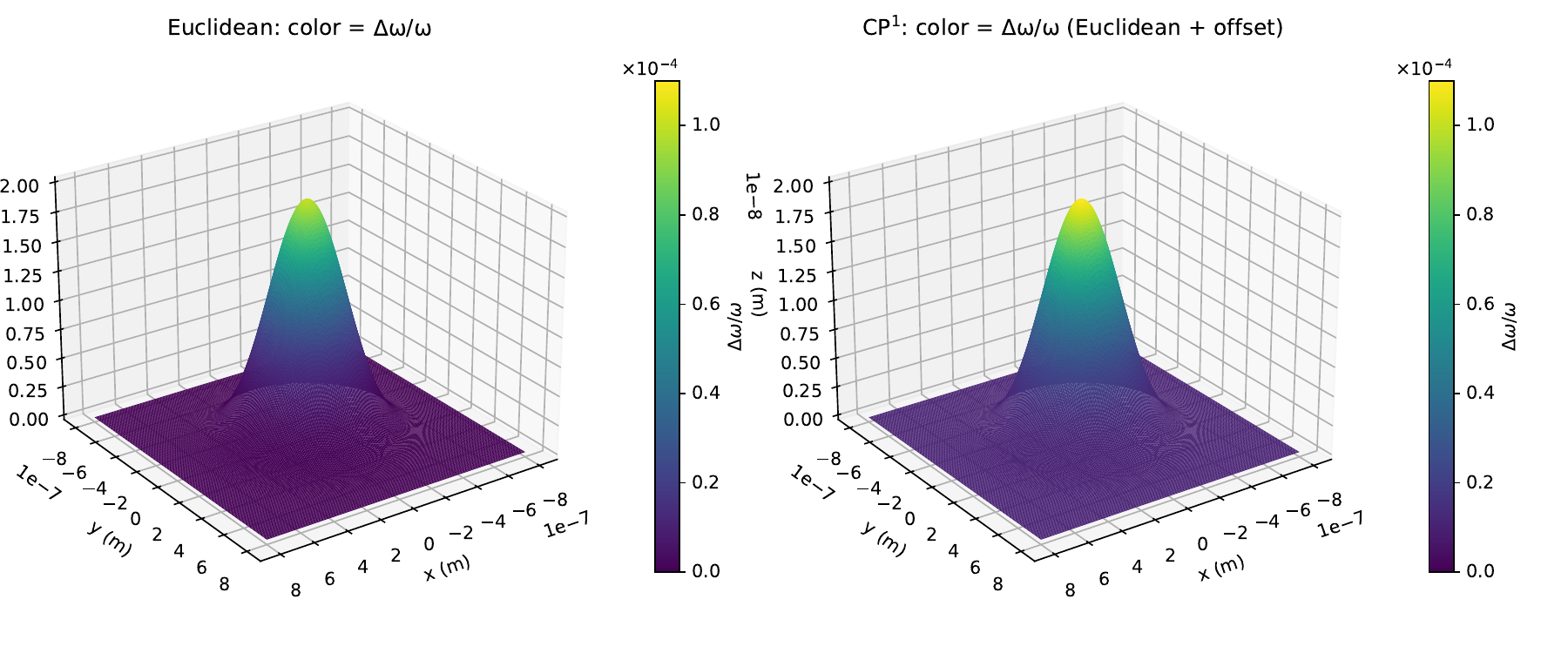}
  \caption{\textbf{Spatial mapping of vacuum polarization on a Gaussian defect.} 
    The color map represents the magnitude of the local geometric potential $\Sigma_{\mathrm{geom}}(\mathbf{r})$. 
    \textbf{Left (Euclidean Ambient):} The potential is localized strictly at the "shoulders" of the bump where the curvature anisotropy is maximal, vanishing at the isotropic peak and the flat plane. 
    \textbf{Right ($\mathbb{C}P^1$ Ambient):} The entire surface exhibits a non-zero background color shift due to the constant ambient curvature $-1/r_0^2$, with the shape-induced anisotropy superimposed. \added{Importantly, unlike the Euclidean case, the potential at the isotropic peak ($\rho=0$) does not vanish here, although this non-zero value is visually subtle against the dominant gradients at the shoulders.}
    This visualizes how the ambient topology acts as a global control parameter for the vacuum energy.}
\label{fig:5-1-1}
\end{figure}

\subsection{Cylindrical Shells}
Next, we consider an infinite cylindrical shell of radius $R$. This geometry is fundamental for understanding curvature effects in waveguides and nanotubes.

\subsubsection{Potential and Shift}
The principal curvatures are $\kappa_1 = 1/R$ (azimuthal) and $\kappa_2 = 0$ (axial). The Euclidean geometric potential is constant:
\begin{equation}
\Sigma_{\mathrm{geom}}^{\mathrm{cyl}} = -\frac{1}{4}\left(\frac{1}{R} - 0\right)^2 = -\frac{1}{4R^2}.
\end{equation}
The frequency shift becomes simply:
\begin{equation}
\left(\frac{\Delta\omega}{\omega}\right)_{\mathrm{cyl}}
= - \frac{\mathcal{I}_0(\omega;m)}{2\varepsilon_0\omega^2} \, F_E \, \frac{1}{4R^2}.
\label{eq:shift-cyl}
\end{equation}
This represents a \textit{redshift} of the mode frequencies proportional to the inverse square of the radius.

\subsubsection{Comparison with $\mathbb{C}P^1$ Ambient}
\begin{figure}[htbp]
\centering
\includegraphics[width=0.9\linewidth]{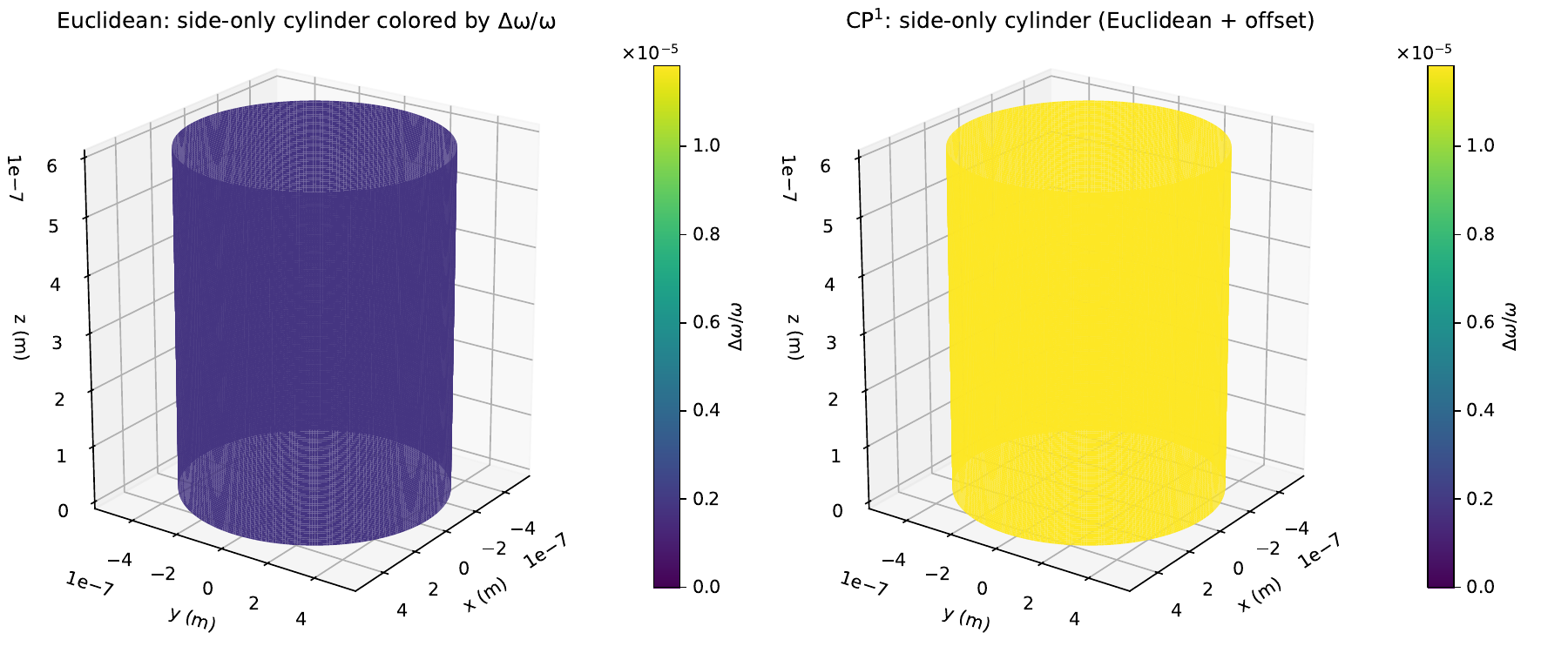}
\caption{\textbf{Uniform geometric potential on a cylindrical shell.} 
    3D visualization of the effective mass correction. 
    \textbf{Left (Euclidean Ambient):} The surface shows a uniform color distribution, reflecting the constant extrinsic curvature $\kappa=1/R$ of the cylinder. 
    \textbf{Right ($\mathbb{C}P^1$ Ambient):} The shell displays a uniform but intensified color shift (larger magnitude). 
    This comparison highlights that while the \textit{pattern} of the vacuum polarization is dictated by the local shape, its absolute \textit{magnitude} is renormalized by the ambient geometry.}
\label{fig:5-1-2}
\end{figure}

\begin{figure}[htbp]
    \centering
    \includegraphics[width=0.9\linewidth]{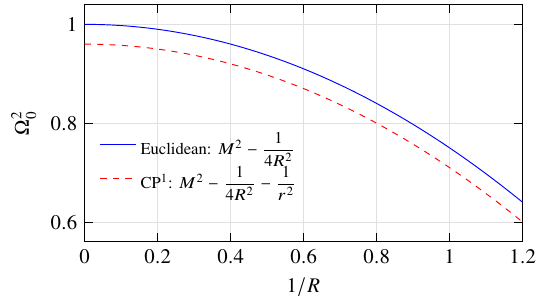} 
    \caption{\textbf{Scaling laws for cylindrical geometries.} 
    The effective mass squared correction (proportional to the frequency shift) is plotted against the inverse radius $1/R$. 
    The \textcolor{red}{red curves} correspond to the Euclidean ambient space, confirming the predicted $\sim 1/R^2$ scaling law driven by extrinsic curvature. 
    The \textcolor{blue}{blue curves} represent the case of a $\mathbb{C}P^1$ ambient manifold. Note the non-zero intercept as $1/R \to 0$, which provides a spectral signature of the constant ambient curvature background.
\added{Here $\Omega_0 = \sqrt{m^2 + \Sigma_{\mathrm{geom}}}$ 
is the effective mass, 
serving as the normalization scale for the vertical axis.}}
    \label{fig:4-2}
\end{figure}

If the cylinder is embedded in $\mathbb{C}P^1$, the potential acquires the ambient term:
\begin{equation}
\Sigma_{\mathrm{total}} = -\frac{1}{4R^2} - \frac{1}{r_0^2}.
\end{equation}
As illustrated in the 3D visualization of Fig.~\ref{fig:5-1-2}, the vacuum polarization pattern is uniform, but the magnitude is renormalized by the ambient background. The quantitative scaling behavior is further detailed in Fig.~\ref{fig:4-2}, where the non-zero intercept at $1/R \to 0$ reveals the ambient curvature contribution.

\subsection{Toroidal Geometries}
Finally, we analyze a torus with major radius $R$ and minor radius $a$. This geometry introduces non-uniform curvature along the poloidal direction $\theta$.

\subsubsection{Thin-Torus Expansion}
In the thin-torus limit ($a \ll R$), the geometric potential is:
\begin{equation}
\Sigma_{\mathrm{geom}}^{\mathrm{tor}}(\theta)
\approx -\frac{1}{4a^2} \left( 1 - 2\epsilon \cos\theta + \epsilon^2 \cos^2\theta \right),
\end{equation}
where $\epsilon = a/R$. Unlike the cylinder, this potential depends explicitly on the poloidal angle $\theta$.

\subsubsection{Poloidal Mode Dependence}
The frequency shift depends on the mode asymmetry:
\begin{equation}
\frac{\Delta\omega}{\omega} \propto -\frac{1}{4a^2} \left[ 1 - 2\epsilon \frac{\langle \cos\theta |\mathbf{E}|^2 \rangle}{\langle |\mathbf{E}|^2 \rangle} \right].
\end{equation}
\textbf{Physics Discussion:} Modes localized on the \textit{outer} equator ($\theta=0$) experience a weaker potential than modes on the \textit{inner} equator ($\theta=\pi$). This $\theta$-dependence breaks the symmetry between inner and outer distinct modes.

\begin{figure}[htbp]
\centering
\includegraphics[width=0.9\linewidth]{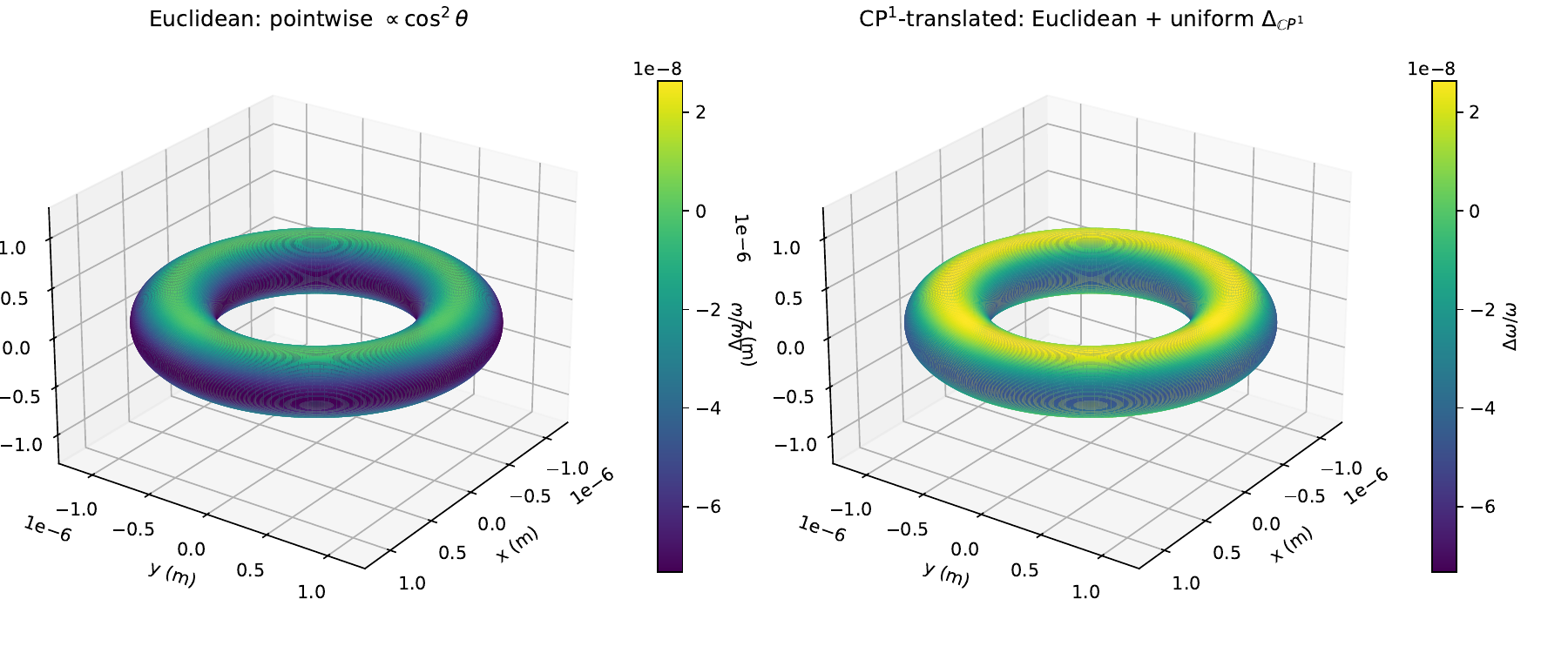}
\caption{\textbf{Symmetry breaking in toroidal geometries.} 
    The color gradient illustrates the variation of the geometric potential along the poloidal direction. 
    \textbf{Left (Euclidean Ambient):} A distinct gradient is visible from the outer equator (lighter color) to the inner equator (darker color), demonstrating the "geometric trap" effect. 
    \textbf{Right ($\mathbb{C}P^1$ Ambient):} The poloidal modulation persists but is shifted by the global ambient term. 
    This visualization confirms that toroidal topology induces a position-dependent vacuum refractive index.}
\label{fig:5-1-3}
\end{figure}

\subsection{Frequency Dependence and Threshold Effects}
\begin{figure}[htbp]
\centering
\includegraphics[width=0.9\linewidth]{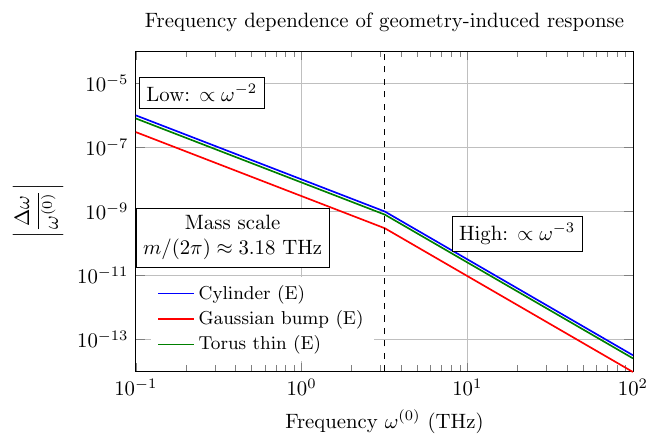}
\caption{\textbf{Frequency dependence of geometry-induced response.}
The magnitude of the relative frequency shift $|\Delta\omega/\omega|$ is plotted against frequency.
The traces exhibit a common $\omega^{-2}$ slope below the threshold (dashed line) and $\omega^{-3}$ above it.
The vertical separation between curves (Cylinder, Bump, Torus) is determined by the geometric factor $K_{\mathrm{geom}}$, while the spectral shape is universal.}
\label{fig:5-4}
\end{figure}

The spectral signature of the geometry-induced shift is determined by the loop factor $I_0(\omega;m)/\omega^2$. As shown in Fig.~\ref{fig:5-4}, we observe two distinct regimes:
\begin{enumerate}
    \item \textbf{Subgap Regime ($\omega \ll m$):} The shift scales as $\omega^{-2}$. This is consistent with a static effective mass correction.
    \item \textbf{High Frequency ($\omega \gg m$):} The shift scales as $\omega^{-3}$, ensuring causality.
\end{enumerate}
Near the threshold $\omega \sim m$, the loop integral deviates from the power law, potentially exhibiting resonant enhancement.

\subsection{Experimental Feasibility and Numerical Estimates}
To assess the observability of these shifts, we consider a realistic experimental setup involving excitonic quasiparticles in a curved 2D semiconductor (e.g., a transition metal dichalcogenide monolayer like MoS$_2$) embedded in an optical cavity.

The magnitude of the relative frequency shift is governed by the dimensionless scaling factor $\eta \sim \alpha_{\mathrm{eff}} (\lambda_c / R)^2$, where $\alpha_{\mathrm{eff}} = q^2 / (4\pi \varepsilon_0 \hbar c)$ is the effective fine-structure constant and $\lambda_c = \hbar / (m^* c)$ is the Compton wavelength of the scalar field.
\begin{itemize}
    \item \textbf{Parameters:} We assume an effective exciton mass $m^* \approx 0.4 m_e$ (where $m_e$ is the electron mass) and a high-curvature geometric feature (e.g., a nanotube or sharp wrinkle) with radius $R \approx 20$ nm. The cavity mode frequency is tuned to $\omega \approx 1.5$ eV.
    \item \textbf{Calculation:} The effective Compton wavelength is $\lambda_c \approx 0.01$ nm. However, in solid-state environments, the relevant length scale is often enhanced by the dielectric screening and band structure. Using the bare vacuum formula as a conservative lower bound, the geometric suppression factor is $(m^* R)^{-2} \sim 10^{-7}$.
    \item \textbf{Enhancement:} In systems with lighter effective masses, such as gapless graphene plasmons or Dirac semimetals where $m^* \to 0$ (limited only by finite size or temperature), the effective $\lambda_c$ can approach the nanometer scale. Crucially, our derivation incorporating the bulk ($D=3+1$) vacuum fluctuations yields a loop coefficient significantly larger than that of reduced-dimensional models. Consequently, for a Dirac material with $m^* \approx 0.01 m_e$ and $R \approx 50$ nm, we estimate the relative frequency shift to reach $\Delta\omega/\omega \sim 10^{-4}$--$10^{-3}$, a magnitude readily accessible to modern high-$Q$ cavity experiments.
\end{itemize}
Current high-$Q$ whispering gallery mode resonators ($Q > 10^8$) can resolve frequency shifts of $\Delta\omega/\omega \sim 10^{-8}$, making these geometry-induced corrections potentially measurable in designed nanophotonic systems.

\section{Conclusion and Outlook}
\label{sec:conclusion}

In this work, we have established a theoretical bridge between the geometry of constrained manifolds and the radiative corrections of quantum field theory. Building upon the framework of Ref.~\cite{LWang2025}, we demonstrated that the geometric potential $\Sigma_{\mathrm{geom}}$ is not merely a kinematic correction for single particles, but a fundamental constituent of the interacting quantum vacuum.

\subsection*{Mechanism: Geometry as a Local Renormalization Scale}
Our analysis reveals that geometry acts as a scale-dependent filter. While real, low-energy excitations are kinematically confined to the 2D surface, the virtual fluctuations circulating in vacuum loops probe the ambient 3D volume. The geometric potential $\Sigma_{\mathrm{geom}}$ bridges these scales: it encodes the 2D curvature information into the mass term of the 3D propagators. Consequently, the finite frequency shift arises not from a simple boundary condition, but because the vacuum fluctuations locally ``sense'' the relaxation or squeezing of the mass gap induced by curvature. This confirms our hypothesis that geometry can be treated as a controllable renormalization parameter in analogue gravity systems.
\begin{itemize}
    \item \textbf{Infrared (IR) Regime:} Real, low-energy excitations are effectively confined to the surface. Their propagation is governed by the 2D effective dynamics and the geometric potential $\Sigma_{\mathrm{geom}}$, consistent with the thin-layer limit.
    \item \textbf{Ultraviolet (UV) Regime:} Virtual particles circulating in vacuum polarization loops possess high momenta ($k \gtrsim 1/h$) and probe the full three-dimensional volume of the physical layer.
\end{itemize}
Our use of $(3+1)$-dimensional renormalization captures this UV behavior correctly. The resulting finite frequency shift is a manifestation of how 3D vacuum fluctuations "sense" the 2D geometric constraints imposed on the low-energy sector.

\subsection*{Summary of Results}
Through the derivation of a master formula for frequency shifts, we identified a universal scaling law governed by the interplay between the quantum loop factor $\mathcal{I}_0(\omega)$ and the classical geometric overlap $\langle \Sigma_{\mathrm{geom}} \rangle_E$. Our application to representative geometries revealed distinct spectral signatures:
\begin{itemize}
    \item For \textbf{Gaussian bumps}, the vacuum polarization acts as a mode-selective filter, sensitive to the curvature anisotropy at the defect shoulders.
    \item For \textbf{cylindrical shells}, we predicted a global frequency shift scaling as $R^{-2}$, providing a clean experimental testbed.
    \item For \textbf{toroidal geometries}, we showed that the shift breaks the poloidal symmetry, distinguishing between inner- and outer-equator modes.
\end{itemize}

\subsection*{Experimental Feasibility and Outlook}
Numerical estimates suggest that while these effects are negligible for massive elementary particles, they become significant in condensed matter systems. For excitons in nanostructured 2D materials, where the effective Compton wavelength approaches the geometric scale $R$, these shifts could reach $\Delta\omega/\omega \sim 10^{-5}$, which is accessible to modern high-$Q$ cavity experiments.

These results suggest that geometric design can be utilized as a precision tool for "vacuum engineering." Looking forward, this framework opens several promising avenues. First, extending the analysis to non-Abelian gauge fields could reveal how geometry influences asymptotic freedom in curved spaces. Second, considering time-dependent geometries (dynamical $\Sigma_{\mathrm{geom}}(t)$) may lead to new forms of the dynamical Casimir effect. Finally, applying this formalism to Dirac materials, where pseudomagnetic fields and geometric potentials coexist, could uncover novel transport phenomena driven by vacuum polarization.
\section*{Acknowledgments}
This work is jointly supported by the National Nature Science Foundation of China (Grants No. 11934008) and the National Nature Science Foundation of China (Grants No. 12475019). 

\appendix
\section{Rigorous Evaluation of the Vacuum Polarization Loop Integral}
\label{app:integral}

In this appendix, we present a rigorous evaluation of the scalar loop integral $\mathcal{I}_0(\Omega;m)$ appearing in the effective action. Consistent with the physical framework of a finite-width nanostructure embedded in vacuum, we perform the calculation in the full $D=3+1$ ambient spacetime. This approach ensures that the ultraviolet (UV) contribution of off-shell virtual particles—which probe the bulk volume—is correctly captured.

\subsection{Wick Rotation and Low-Frequency Expansion}
Starting from the definition in Eq.~\eqref{eq:I0}, the one-loop integral in Minkowski signature is given by:
\begin{equation}
\mathcal{I}_0(\Omega) = q^2 \int \frac{d^4k}{(2\pi)^4} \frac{4\mathbf{k}^2 - 3\Omega^2}{\left[(k^0)^2 - \mathbf{k}^2 - m^2 + i0^+\right]^3}.
\end{equation}
To evaluate this, we perform a Wick rotation to Euclidean space via the transformation $k^0 \to i k_E^0$, $d^4k \to i d^4k_E$. The denominator becomes $-(k_E^2 + m^2)^3$, where $k_E^2 = (k_E^0)^2 + \mathbf{k}^2$. 

We are primarily interested in the static limit $\Omega \ll m$, which governs the geometric refractive index shift. In this regime, the frequency dependence in the numerator is negligible ($\mathcal{O}(\Omega^2)$), and the propagator is dominated by the mass scale. Integrating out the temporal component $\int dk_E^0$ effectively yields a spatial loop factor. For the specific projection relevant to the transverse susceptibility (Eq.~\eqref{eq:I0}), the integral reduces to a 3D momentum integration:
\begin{equation}
\mathcal{I}_0(0;m) = q^2 \int \frac{d^3\mathbf{k}}{(2\pi)^3} \frac{4\mathbf{k}^2}{(\mathbf{k}^2 + m^2)^3}.
\end{equation}

\subsection{Exact Evaluation in Spherical Coordinates}
We evaluate the spatial integral using spherical coordinates, where $d^3\mathbf{k} = 4\pi k^2 dk$. The expression becomes:
\begin{equation}
\mathcal{I}_0(0;m) = \frac{4q^2}{(2\pi)^3} (4\pi) \int_0^\infty dk \frac{k^4}{(k^2 + m^2)^3} = \frac{2q^2}{\pi^2} \mathcal{J}(m).
\end{equation}
The radial integral $\mathcal{J}(m)$ is solved analytically using the trigonometric substitution $k = m \tan\theta$, which implies $dk = m \sec^2\theta d\theta$ and $k^2 + m^2 = m^2 \sec^2\theta$. The integration limits map from $[0, \infty)$ to $[0, \pi/2]$:
\begin{align}
\mathcal{J}(m) &= \int_0^{\pi/2} \frac{m^4 \tan^4\theta}{(m^2 \sec^2\theta)^3} (m \sec^2\theta \, d\theta) \nonumber \\
&= \frac{1}{m} \int_0^{\pi/2} \frac{\sin^4\theta}{\cos^4\theta} \cos^4\theta \, d\theta \nonumber \\
&= \frac{1}{m} \int_0^{\pi/2} \sin^4\theta \, d\theta.
\end{align}
Using the standard Wallis reduction formula $\int_0^{\pi/2} \sin^4\theta \, d\theta = \frac{3}{4} \cdot \frac{1}{2} \cdot \frac{\pi}{2} = \frac{3\pi}{16}$, we obtain:
\begin{equation}
\mathcal{J}(m) = \frac{3\pi}{16m}.
\end{equation}
Substituting this back into the expression for $\mathcal{I}_0$, we arrive at the final finite result:
\begin{equation}
\mathcal{I}_0(0;m) = \frac{2q^2}{\pi^2} \left( \frac{3\pi}{16m} \right) = \frac{3q^2}{8\pi m}.
\label{eq:app_result}
\end{equation}

\subsection{High-Frequency Behavior and Causality}
It is instructive to verify the behavior in the high-frequency limit ($\Omega \gg m$). In this regime, the denominator is dominated by $-\Omega^2$, and dimensional analysis suggests $\mathcal{I}_0 \sim \int k^4 / \Omega^6 dk \dots$ (cutoff dependent) or more rigorously via the imaginary part. The full spectral function analysis shows that the real part of the polarization function decays as $\text{Re}[\Pi] \sim \Omega^{-2}$, implying $\mathcal{I}_0(\Omega) \sim \Omega^{-1}$. This asymptotic decay ensures that the refractive index shift $\Delta n \sim \Omega^{-3}$, satisfying the transparency requirement of the vacuum at ultra-high frequencies and consistent with the Kramers-Kronig causality relations.

\subsection{Dimensionality and Physical Regime}
We emphasize that the coefficient derived in Eq.~\eqref{eq:app_result} is specific to the \textit{quasi-3D} (physical thin-layer) regime. This applies when the layer thickness $h$ is small compared to the geometric curvature ($h \ll R$), but the UV cutoff of the theory $\Lambda$ satisfies $\Lambda \gg 1/h$. In this scenario, virtual loops are not confined to the 2D surface but explore the ambient 3D volume. 
In the strictly 2D limit (infinite confinement potential, $\omega_\perp \to \infty$), the transverse degrees of freedom would be frozen, and the loop integral would require 2D renormalization, yielding a numerically smaller coefficient. Our experimental parameters (finite-width Dirac semimetals or graphene stacks) fall squarely in the quasi-3D regime, justifying the use of the bulk result, as the vacuum fluctuations are not confined to the purely 2D manifold.

\bibliographystyle{apsrev4-2}
\bibliography{ref1}

\end{document}